\documentclass[11pt]{article}
\usepackage{amsmath}
\usepackage{times}
\usepackage{graphicx}
\usepackage{color}
\usepackage{multirow}
\usepackage{lineno}
\usepackage{algorithm,caption}
\usepackage[noend]{algpseudocode}
\usepackage{amssymb}
\usepackage[onehalfspacing]{setspace}
\usepackage{changepage}
\usepackage[round]{natbib}
\usepackage{bbm}
\usepackage{latexsym}

\title{
  Eligibility Traces and Plasticity on Behavioral Time Scales: Experimental Support of neoHebbian Three-Factor Learning Rules}

\author{Wulfram Gerstner$^{*}$, Marco Lehmann, Vasiliki Liakoni, Dane Corneil, and Johanni Brea\\
	School of Computer Science \& School of Life Sciences\\
	\'Ecole Polytechnique F\'ed\'erale de Lausanne,\\
	1015 Lausanne EPFL, Switzerland\\
(*) corresponding author}

\begin{document}

\maketitle

\begin{abstract}

  Most elementary behaviors such as moving the arm to grasp an object or walking into the next room to explore a museum  evolve on the time scale of seconds;
  in contrast, neuronal action potentials occur on the time scale of a few milliseconds.  Learning rules of the brain must therefore bridge the gap between these two different time scales.
  Modern theories of synaptic plasticity have postulated that the co-activation of pre- and postsynaptic neurons sets a flag at the synapse, called an eligibility trace, that leads to a weight change only if an additional factor is present while the flag is set. This third factor, signaling reward, punishment, surprise, or novelty, could be implemented by the phasic activity of neuromodulators or specific neuronal inputs signaling special events. While the theoretical framework has been developed over the last decades, experimental evidence in support of eligibility traces on the time scale of seconds has been collected only during the last few years.
  Here we review, in the context of three-factor rules of synaptic plasticity,  four key experiments that support the role of synaptic eligibility traces in combination with a third factor as a biological implementation of neoHebbian three-factor learning rules.
  
\end{abstract}


\section{Introduction}

Humans are able to learn novel behaviors such as pressing a button, swinging a tennis racket, or breaking at a red traffic light; they are also able to form memories of salient events, learn to distinguish flowers,
and to  establish a mental map when exploring a novel environment.
Memory formation and behavioral learning is linked to changes of synaptic connections \citep{Martin00}.
Long-lasting synaptic changes, necessary for memory,  can be induced by Hebbian protocols that combine the activation of presynaptic terminals with a manipulation of the voltage or the firing state of the postsynaptic neuron \citep{Lisman03}.
Traditional experimental protocols of long-term potentiation (LTP) \citep{Bliss73a,Bliss93}, long-term depression (LTD) \citep{Levy83,Artola93}  and spike-timing dependent plasticity (STDP)
\citep{Markram97,Zhang98,Sjostrom01}
neglect that additional factors such as neuromodulators or other gating signals  might be necessary to
permit synaptic changes \citep{Gu02,Hasselmo06,Reynolds02}.
Early STDP  experiments that involved neuromodulators mainly focused on tonic bath application of modulatory factors
\citep{Pawlak10}.
However, from the perspective of formal learning theories, to be reviewed below, the timing of modulatory factors is just as crucial \citep{Schultz00,Schultz02}.
From the theoretical perspective, STDP under the control of neuromodulators leads to the framework of three-factor learning rules \citep{Xie04,Legenstein08b,Vasilaki09} where an eligibility trace represents the Hebbian idea of co-activation of pre- and postsynaptic neurons \citep{Hebb49} while modulation of plasticity by additional
gating signals is represented generically by a 'third factor' \citep{Crow68,Barto85,Legenstein08b}.
Such a third factor could represent variables such as 
'reward minus expected reward' \citep{Williams92,Schultz98,Sutton98}
or the saliency of an unexpected event
\citep{Ljunberg92,Redgrave06}.

In an earlier paper \citep{Fremaux16}
we reviewed the theoretical literature of,  and experimental support for, three-factor rules available by the end of 2013.
During recent years, however, the experimental procedures advanced significantly and provided direct physiological evidence of eligibility traces and three-factor learning rules for the first time,
making an updated review of three-factor rules necessary.
In the following, we -- a group of theoreticians -- review 
five experimental papers indicating support of eligibility traces in
striatum \citep{Yagishita14},
cortex \citep{He15},
and hippocampus \citep{Brzosko15,Brzosko17,Bittner17}.
We will close with a few remarks
on the paradoxical nature of theoretical predictions in the field of computational neuroscience.

\section{Hebbian rules versus three-factor rules}

Learning rules describe the change of the strength of a synaptic contact between a presynaptic neuron $j$ and a postsynaptic neuron $i$.
The strength of an excitatory synaptic contact can be defined by the amplitude of the postsynaptic potential which is closely related to the spine volume and the number of AMPA
receptors \citep{Matsuzaki01}.
Synapses contain complex molecular machineries \citep{Lisman03,Redondo11,Huganir13,Lisman17},
but for the sake of transparency of the arguments,  we will keep the mathematical notation as simple as possible and  characterize the synapse by two variables only:
the first one is the synaptic 
 strength $w_{ij}$, measured as spine volume or amplitude of postsynaptic potential,
and the second one is a synapse-internal variable $e_{ij}$ which is not directly visible in standard electrophysiological experiments.
In our view, the internal variable $e_{ij}$ represents a metastable transient state of interacting molecules in the spine head
or a multi-molecular substructure in the postsynaptic density which serves as a synaptic flag indicating that the synapse is ready for an increase or decrease of its spine volume
\citep{Bosch14}. 
The precise biological nature of $e_{ij}$ is not important to understand the theories and experiments that are reviewed below.
We refer to $e_{ij}$ as the 'synaptic flag' or the 'eligibility trace'
and to $w_{ij}$ as the 'synaptic weight' or 'strength' of the synaptic contact.
A change of the synaptic flag indicates a 'candidate weight change' \citep{Fremaux10}
whereas a change of $w_{ij}$ indicates an actual, measurable, change of the synaptic weight.
Before we turn to three-factor rules, let us discuss conventional models of Hebbian learning.

\subsection{Hebbian learning rules}

Hebbian learning rules are the mathematical summary of the outcome of experimental protocols
inducing long-term potentiation (LTP) or long-term depression (LTD) of synapses.
Suitable experimental protocols include strong extracellular stimulation of presynaptic fibers \citep{Bliss73a,Levy83},
manipulation of postsynaptic voltage in the presence of presynaptic spike arrivals \citep{Artola93},
or spike-timing dependent plasticity (STDP) \citep{Markram97,Sjostrom01}.
In all mathematical formulations of Hebbian learning, the synaptic flag variable $e_{ij}$ is 
sensitive to the {\em combination} of presynaptic spike arrival and a postsynaptic variable,
such as the voltage
at the location of the synapse.
Under a Hebbian learning rule,
repeated presynaptic spike arrivals at a synapse of a neuron at rest do not
cause a change of the synaptic variable.
Similarly, an elevated postsynaptic potential in the absence of a presynaptic spike  does not cause a change
of the synaptic variable.
Thus Hebbian learning always needs two factors for a synaptic change: a factor caused by a presynaptic signal such as glutamate;
and a factor that depends on the state of the postsynaptic neuron.

What are these factors? We can think of the presynaptic factor
as the time course of glutamate available in the synaptic cleft or bound to the postsynaptic membrane.
Note that the term 'presynaptic factor' that we will use in the following
{\em does not imply}  that the physical location of the presynaptic factor  is inside the presynaptic terminal - the factor could very well be located in the postsynaptic membrane as long as it  {\em only} depends on the amount of available neurotransmitters.
The postsynaptic factor might be the calcium in the synaptic spine
\citep{Shouval02,Rubin05},
a calcium-related second messenger molecule \citep{Graupner07},
or simply the voltage at the site of the synapse \citep{Brader07,Clopath10}.

We remind the reader that we always use the index $j$ to refer to the presynaptic neuron and the index $i$ to refer to the postsynaptic one.
For the sake of simplicity, let us call the presynaptic factor $x_j$
(representing the activity of the presynaptic neuron or the amount of glutamate in the synaptic cleft) and the postsynaptic factor $y_i$ (representing the state of the postsynaptic neuron).
In a Hebbian learning rule, changes of the synaptic flag $e_{ij}$
need both $x_j$ and $y_i$
\begin{equation} \label{eligibility-1}
  \frac{d}{dt} e_{ij} =  \eta \, x_j \, g(y_i) - e_{ij}/\tau_e
\end{equation}
where $\eta$ is the constant learning rate, $\tau_e$ is a decay time constant
and $g(y_j)$ is some arbitrary, potentially nonlinear, function of the postsynaptic variable $y_i$.
Thus, the synaptic flag $e_{ij}$ acts as a {\em correlation detector} between presynaptic activity $x_j$ and the state of the postsynaptic neuron $y_i$. In some models, there is no decay or the decay is considered negligible on the time scale
of one experiment ($\tau_e \to \infty$).

\begin{figure}[tp]
  \hbox{ {\bf A}\hspace{7cm} {\bf B}}
\hspace{5mm}
  \includegraphics[width=6.5cm]{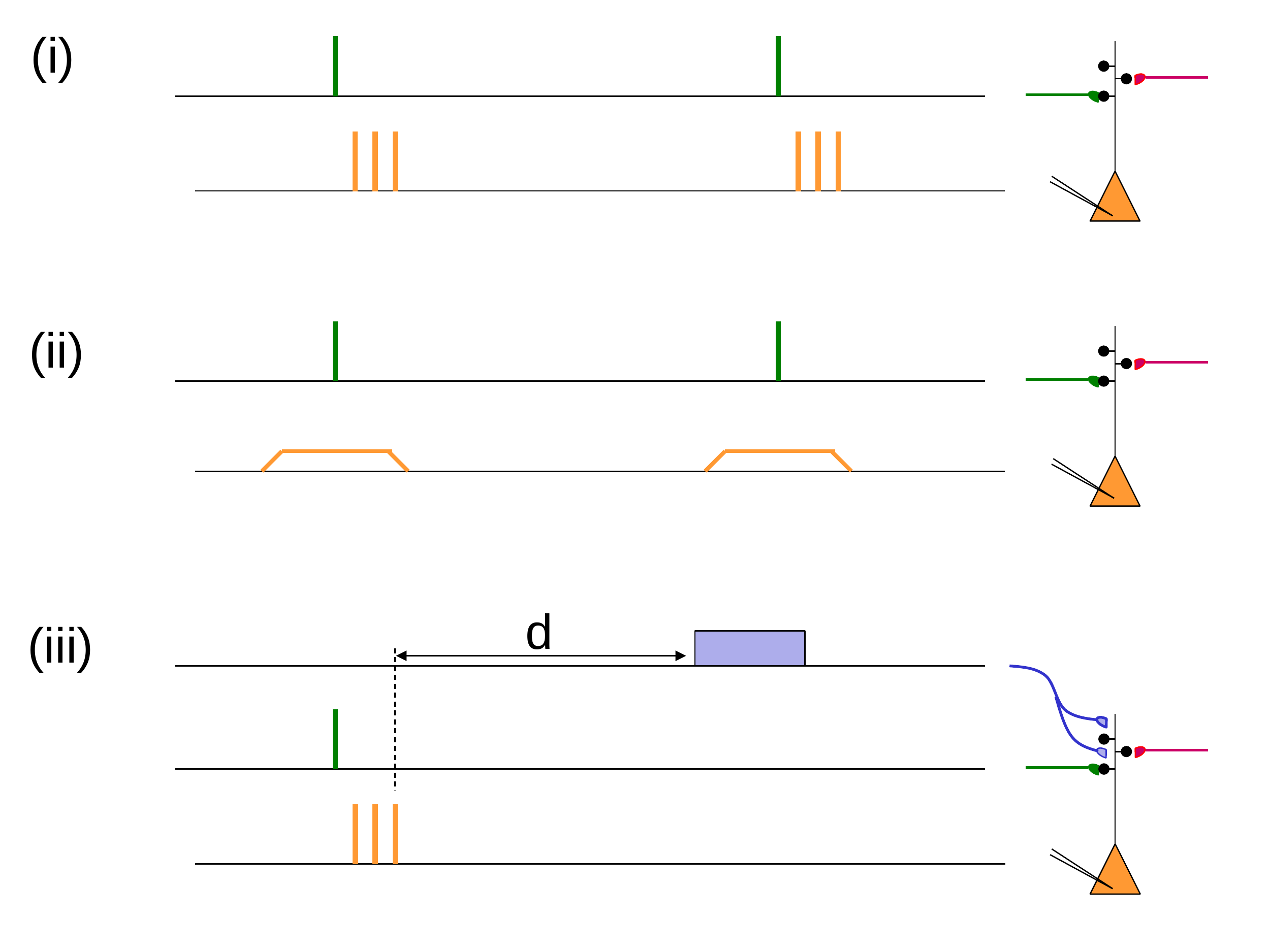}
  \hspace{5mm}
  \includegraphics[width=6.5cm]{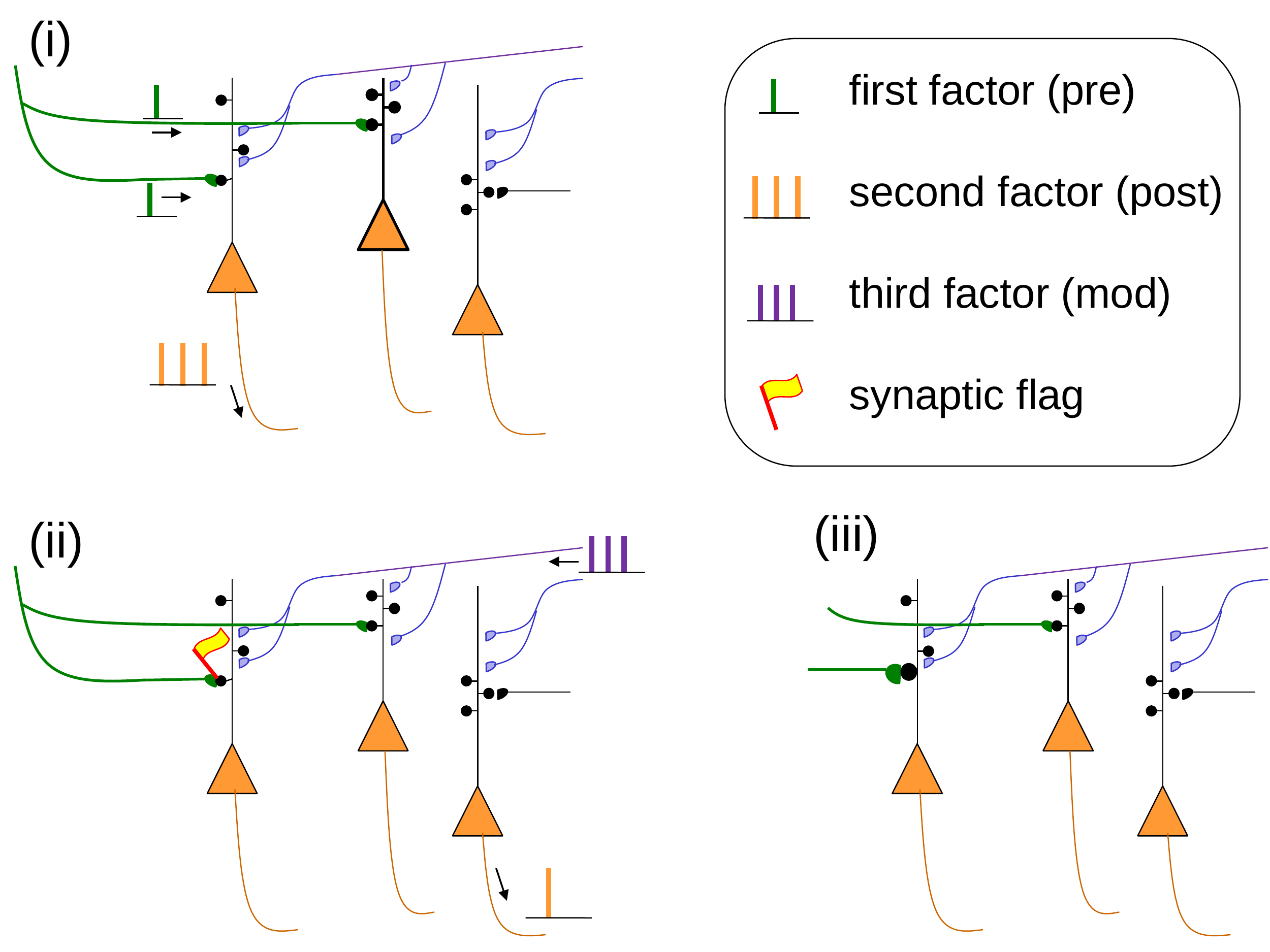}
\caption{\label{fig-1}
  {\bf A.}
  Two Hebbian protocols and one three-factor learning protocol.
    (i) Hebbian STDP protocol with
    presynaptic spikes (presynaptic factor) followed by a burst of postsynaptic spikes (postsynaptic factor).
    Synapses in the stimulated pathway (green) will typically show LTP while an unstimulated synapse (red) will not change its weight \citep{Markram97}.
    (ii) Hebbian voltage pairing protocol of presynaptic spikes (presynaptic factor)
    with a depolarization of the postsynaptic neuron (postsynaptic factor).
    Depending on the amount of depolarization the stimulated pathway (green) will show LTP or LTD while an unstimulated synapse (red) does not change its weight \citep{Artola93,Ngezahayo00}.
    (iii) Results of a Hebbian induction protocol are influenced by a third factor (blue) even if it is given after a delay $d$. The third factor could be a neuromodulator such as dopamine, acetylcholine, noreprinephrine, or serotonin
    \citep{Pawlak10,Yagishita14,He15,Brzosko15,Brzosko17,Bittner17}.
    {\bf B.}
    Specificity of three-factor learning rules. (i) Presynaptic input spikes (green) arrive at two different neurons, but only one of these also shows postsynaptic activity (orange spikes).
   (ii) A synaptic flag is set only at the synapse with a Hebbian co-activation of pre- and postsynaptic factors; the synapse become then eligible to interact with the third factor (blue).  Spontaneous spikes of other neurons do not interfere.
   (iii) The interaction of the synaptic flag with the third factor leads to a strengthening of the synapse (green). 
  }
  \end{figure}

Let us discuss two examples. In the Bienenstock-Cooper Munro (BCM)  model of developmental cortical plasticity \citep{Bienenstock82}
the presynaptic factor $x_j$ is the firing rate of the presynaptic neuron and $g(y_i)=(y_i-\theta)\,y_i$
is a quadratic function with $y_i$ the postsynaptic firing rate and $\theta$ a threshold rate.
Thus, if both pre- and postsynaptic neurons fire together at a high rate $x_j=y_i>\theta$ then
the synaptic flag $e_{ij}$ increases. 
In the BCM model, just like in most other conventional models,
a change of the synaptic flag (i.e., an internal state of the synapse)
leads instantaneously to a change of the weight $e_{ij}\longrightarrow w_{ij}$
so that an experimental protocol results immediately in a measurable weight change.
With the BCM rule and other similar rules \citep{Oja82,Miller94}, the synaptic weight increases
if both presynaptic and postsynaptic neuron are highly active, implementing the slogan 'fire together, wire together' \citep{Lowel92,Shatz92}; cf. Fig.~\ref{fig-1}A(i).

As a second example, we consider the Clopath model \citep{Clopath10}. In this model,
there are two correlation detectors implemented as synaptic flags
$e_{ij}^+$
and $e_{ij}^-$ for LTP and LTD, respectively. 
The synaptic flag $e_{ij}^+$
for LTP
uses a presynaptic factor $x_j^+$ 
(related to
the amount of glutamate available in the synaptic cleft)
which 
increases with each presynaptic spike and decays back to zero over the time of a few milliseconds \citep{Clopath10}.
The postsynaptic factor for LTP depends on the postsynaptic voltage $y_i$ via a 
function $g(y_i) = a_+ [y_i-\theta_+] \bar{y_i}$ where $a_+$ is a positive constant,
$\theta_+$ a voltage threshold, square brackets denote the rectifying piecewise linear function,
and $\bar{y_i}$ a running average of the voltage with a time constant of tens of milliseconds.
An analogous, but simpler, combination of presynaptic spikes and postsynaptic voltage defines the second
synaptic flag $e_{ij}^-$ for LTD \citep{Clopath10}.
The total change of the synaptic weight is the combination of
the two synaptic flags for LTP and LTD:
$dw_{ij}/dt = de_{ij}^+/dt  - de_{ij}^-/dt $.
Note that, since  both synaptic flags $e_{ij}^+$ and $e_{ij}^-$
depend on the postsynaptic voltage, postsynaptic spikes are not a necessary condition for changes,
in agreement with voltage-dependent protocols
\citep{Artola93,Ngezahayo00}.
Thus, in voltage-dependent protocols, and similarly in voltage-dependent models,
'wiring together' is possible without 'firing together' - indicating
that the theoretical framework sketched above goes beyond a narrow view of
Hebbian learning; cf. Fig.~\ref{fig-1}A(ii).

If we restrict the discussion of the postsynaptic variable to super-threshold spikes,
then the Clopath model
becomes identical to the triplet STDP model \citep{Pfister06} which is in turn closely related
to other nonlinear STDP models \citep{Senn01,Froemke02,Izhikevich03a}
as well as to the BCM model discussed above \citep{Pfister06,Gjorjieva11}.
Classic pair-based STDP models \citep{Gerstner96,Kempter99c,Song00,Rossum00,Rubin01}
are further examples of the general theoretical framework of Eq. (\ref{eligibility-1})
and so are some  models of structural plasticity \citep{Helias08,Deger12,Fauth15}.
Hebbian models of synaptic consolidation have several  hidden flag variables
\citep{Fusi05,Barrett09,Benna16} but can also be situated as examples
within the general framework of Hebbian rules.
Note that in most of the examples so far the measured synaptic weight is a
linear function of the synaptic flag variable(s).
However,
this does not need to be the case. For example, in some voltage-based \citep{Brader07}
or calcium-based models \citep{Shouval02,Rubin05},
the synaptic flag is transformed into a weight change only if $e_{ij}$ is above or below some threshold, or only after some further filtering.

To summarize, in the theoretical literature the class of Hebbian models
is a rather general framework encompassing all those models that are driven by a combination of presynaptic activity and the state of the postsynaptic neuron. In this view, Hebbian models depend  on two factors related to the activity of the presynaptic and the state of the postsynaptic neuron.
The correlations between the two factors can be extracted on different time scales using one or, if necessary, several flag variables.
The flag variables trigger a change of the measured synaptic weight. 
In the following we build on Hebbian learning, but extend the theoretical framework to include a third factor.

\subsection{Three-factor learning rules}
We are interested in a framework where a Hebbian co-activation of two neurons leaves one or several flags (eligibility trace) at the synapse connecting these neurons. The flag is not directly visible and does not automatically trigger a change of the synaptic weight. An actual weight change is implemented only if a third signal, e.g., a phasic increase of neuromodulator activity or an additional input (signaling the occurrence of a special event) is present at the same time or in the near future.
Theoreticians refer to such a plasticity model as a three-factor learning rule
\citep{Xie04,Legenstein08b,Vasilaki09,Fremaux13,Fremaux16}.
Three-factor rules have also been called 'neoHebbian' \citep{Lisman11,Lisman17}
or 'heterosynaptic (modulatory-input dependent)' \citep{Bailey00}
and can be traced back to the 1960s \citep{Crow68}, if not earlier.
To our knowledge the wording 'three factors' was first used by \citep{Barto85}.
The terms eligibility and  eligibility traces have
been used in \citep{Klopf72,Sutton81,Barto83,Barto85,Williams92,Schultz98,Sutton98} but in
some of the early studies it remained unclear whether eligibility traces can be set by
  presynaptic activity alone
  \citep{Klopf72,Sutton81}
or only by Hebbian co-activation of pre- and postsynaptic neurons \citep{Barto83,Barto85,Williams92,Schultz98,Sutton98}.

The basic idea of a modern eligibility trace
is that a synaptic flag variable $e_{ij}$ is set according to Eq. (\ref{eligibility-1})
by  coincidences between presynaptic activity $x_j$ and a postsynaptic factor $y_i$.
The update of the synaptic weight $w_{ij}$, as measured via the spine volume or the amplitude
of the excitatory postsynaptic potential (EPSP), is given by
\begin{equation}\label{three-factor-2}
  \frac{d}{dt}{w_{ij}} = e_{ij} \, M_{3rd}(t)
  \end{equation}
where $M_{3rd}(t)$
refers to the third factor \citep{Izhikevich07,Legenstein08b,Fremaux13}.
Thus, a third factor is needed to transform the eligibility trace into a weight change;
cf. Fig.~\ref{fig-1}A(iii).
Note that the 
weight change is proportional to $M_{3rd}(t)$.
Thus, the third factor influences the speed of learning. In the absence of the third factor
($M_{3rd}(t)=0$), the synaptic weight is not changed.
We emphasize that  a positive
value of the synaptic flag in combination with a negative value of $M_{3rd}(t)$ leads to a decrease of the weight.
Therefore, the third factor also influences the {\em direction} of change.

What could be the source of such a third factor?
The third factor could be triggered by attentional processes, surprising events, or reward.
Phasic signals of neuromodulators such as dopamine, serotonin, acetylcholine, or noradrenaline
are obvious candidates of a third factor, but potentially not the only ones.  Note that axonal branches of most
dopaminergic, serotonergic, cholinergic, or adrenergic neurons
project broadly onto large regions of cortex so that
a phasic neuromodulator signal arrives at many neurons and synapses in parallel
\citep{Schultz98}.
Since neuromodulatory information is shared by many neurons, 
the variable $M_{3rd}(t)$ of the third factor  has no neuron-specific index (neither $i$ nor $j$) in our mathematical formulation.
Because of its unspecific nature, the theory literature sometimes refers to the third factor as a 'global' broadcasting signal,
even though in practice not every brain region and every synapse is reached by each neuromodulator.

Note that we mathematically define a phasic signal as the deviation from the running average so that $M_{3rd}(t)$ in Eq. (\ref{three-factor-2})
can take positive and negative values.
However, the third factor
could also be biologically implemented by {\em positive} excursions
of the activity using two different
neuromodulators with very low baseline activity.
The activity of the first modulator could indicate
positive values of the third factor and  that of the second modulator negative ones - similar to ON and OFF cells in the retina.
Similarly, the framework of neoHebbian three-factor rules is general enough
to enable biological implementations with separate eligibility traces for LTP and LTD as discussed above in the context of the Clopath model \citep{Clopath10}.

\subsection{Examples and theoretical predictions}

There are several known examples
in the theoretical literature of neoHebbian three-factor rules.
We briefly present three of these and formulate
expectations 
derived from the theoretical framework 
which we would like to compare to experimental results in the next section.

As a first example, we consider the relation of neoHebbian three-factor rules to reward-based learning. 
Temporal Difference (TD) algorithms  such as SARSA($\lambda)$ or TD($\lambda$)
from the  theory of reinforcement learning \citep{Sutton98}
as well as learning rules derived from
policy gradient theories \citep{Williams92}
can be interpreted in neuronal networks in terms of
neoHebbian three-factor learning rules.
The resulting plasticity rules are 
applied  to synapses connecting
'state-neurons' (e.g., place cells coding for the current location of an animal)
to 'action neurons' (e.g., cells initiating an action program such as 'turn left')
\citep{Brown95,Suri99,Arleo00,Foster00,Xie04,Loewenstein06,Florian07,Izhikevich07,Legenstein08b,Vasilaki09,Fremaux13};
for a review, see \citep{Fremaux16}.
The eligibility trace is increased during the joint activation of 'state-neurons' and
'action-neurons' and decays
exponentially thereafter consistent with the framework of Eq. (\ref{eligibility-1}).
The third factor is defined as reward minus expected reward where the exact definition
of expected reward depends on the implementation details.
A long line of research by Wolfram Schultz and colleagues \citep{Schultz97,Schultz98,Schultz00,Schultz02} indicates that phasic increases of the neuromodulator dopamine
have the necessary properties required for a third factor in the theoretical framework of reinforcement learning.

However, despite the rich literature on dopamine  and reward-based learning
accumulated during the last 25 years,
measurements of the decay time constant $\tau_e$ of the eligibility trace $e_{ij}$ in Eq. (\ref{eligibility-1})
have, to our knowledge,  not become available before 2015.
From the mathematical framework of neoHebbian three-factor rules
it is clear that, in the context of action learning,
the time constant of the eligibility trace (i.e., the duration of the synaptic flag)
should roughly match the time span from the initiation of an action to the delivery of reward. 
As an illustration, let us imagine a baby that attempts to grasp her bottle of milk.
The typical duration of one grasping movement is in the range of a second, but potentially
only the third grasping attempt might be successful.
Let us suppose that each grasping movement corresponds to the co-activation of some neurons in the brain.
If the duration of the synaptic flag is much less than a second, the co-activation of pre- and postsynaptic neurons
that sets the synaptic flag (eligibility trace)
cannot be linked to the reward one second later and synapses do not change.
If the duration of the synaptic flag is much longer than a second,
then the two 'wrong' grasping attempts are reinforced nearly as strongly as the third, successful one
which mixes learning of 'wrong' co-activations with the correct ones.
Hence, {\em the existing theory of three-factor learning rules predicts that the synaptic flag (eligibility trace for action learning) should be in the range of a typical elementary action, about 200ms to 2s}; see, for example,
p. 15 of \citep{Schultz98}\footnote{'Learning ... (with dopamine)  on striatal synapses ... requires hypothetical traces of synaptic activity that last until reinforcement occurs and makes those synapses eligible for modification ...'},
p.3 of \citep{Izhikevich07},
\footnote{'(the eligibility trace $c$ ) decays to $c=0$ 
  exponentially with the time constant $\tau_c= 1$s'},
p.3 of \citep{Legenstein08b}\footnote{'the time scale
of the eligibility trace is assumed in this article to be on the order of seconds'},
p. 13327 of \citep{Fremaux10}\footnote{'candidate weight changes $e_{ij}$ decay to zero with a
time constant 	$\tau_e=500$ms. The candidate weight changes $e_{ij}$ are known as
the eligibility trace in reinforcement learning'},
or
p. 13 of \citep{Fremaux13}\footnote{'the time scales of the eligibility traces we propose, (are) on
  the order of hundreds of milliseconds, ..
Direct experimental evidence of eligibility traces still lacks, ...'}.
An eligibility trace of 100ms or  20 seconds would be less useful for learning a typical
elementary 
action
or delayed reward task than an eligibility trace in the range of 200ms to 2s. The expected time scale of the synaptic eligibility trace should roughly match the maximal delay of reinforcers in conditioning experiments \citep{Thorndike11,Pavlov27,Black85}, linking synaptic processes to behavior. For human behavior, delaying a reinforcer by 10 seconds during ongoing actions decreases learning compared to immediate reinforcement \citep{Okouchi09}.

As a second example, we consider situations that go beyond standard reward-based learning.
Even in the absence of reward, a surprising event might trigger a combination of neuromodulators
such as noradrenaline, acetylcholine and dopamine that may act as  third factor for synaptic plasticity.
Imagine a small baby lying in the cradle with an attractive colorful object swinging above him.
He spontaneously makes several arm movements until finally he succeeds, by chance,
to grasp the object.
There is no food reward for this action. However, the fact that he can now turn the object, look at it from different sides,
or put it in his mouth is satisfying because it leads to many novel (and exciting!) stimuli.
The basic idea is that, in such situations, novelty or surprise
acts as a reinforcer even in complete absence of food rewards
\citep{Schmidhuber91,Singh04,Oudeyer07}.
Theoreticians have studied these ideas in the context
of curiosity \citep{Schmidhuber10}, information gain during active exploration \citep{Storck95,Sun11,Schmidhuber06,Sun11,Little13,Friston16}, and
via formal definitions of surprise
\citep{Storck95,Itti09,Schmidhuber10,Shannon48,Friston10,Faraji18}.
Note that surprise is not always linked to active exploration but can also occur
in a passive situation,  e.g. listening to tone beeps or viewing simple stimuli
\citep{Squires76,Kolossa13,Kolossa15,Meyniel16}.
Measurable physiological responses to surprise include pupil dilation \citep{Hess60}
and the P300 component of the electroencephalogram \citep{Squires76}.

If surprise can play a role similar to reward, then surprise-transmitting broadcast signals should speed-up plasticity.
Indeed,  
theories of surprise as well as hierarchical Bayesian models predict
 a faster change of model parameters
 for surprising stimuli than for known ones \citep{Yu05,Nassar10,Mathys11,Mathys14,Faraji18} similar to, but more general than,
 the well-known Kalman filters \citep{Kalman60}.
Since the translation of these abstract models into networks of spiking neurons is still missing,
precise predictions for surprise modulation of plasticity in the form of three-factor rules are not yet available. However, 
if we consider noradrenaline, acetylcholine, and/or dopamine as candidate neuromodulators signaling novelty and surprise,
we expect that these neuromodulators should have a strong effect on plasticity
so as to boost learning of surprising stimuli.
The influence of tonic applications of various neuromodulators on synaptic plasticity
has been shown in numerous studies \citep{Gu02,Hasselmo06,Reynolds02,Pawlak10}.
However, in the context of the above examples, we are interested in phasic neuromodulatory signals.
Phasic signals conveying moments of  surprise
are most useful for learning if they are either synchronous with the stimulus to be learned
(e.g., passive listening or viewing) or arise with a delay corresponding to one exploratory movement (e.g. grasping).
Hence, we predict from these considerations a decay constant $\tau_e$ of the synaptic flag
in the range of 1 second, but with a pronounced effect for synchronous or near-synchronous events.

As our final example, we would like to comment on synaptic consolidation.
The synaptic tagging-and-capture hypothesis \citep{Frey97,Reymann07,Redondo11}
perfectly fits in the framework of three-factor learning rules:
The joint pre- and postsynaptic activity sets the synaptic flag (called 'tag' in the context of consolidation)
which decays back to zero over the time of one hour.
To stabilize synaptic weights beyond one hour an additional factor is needed to trigger
protein synthesis required for long-term maintenance of synaptic weights \citep{Redondo11,Reymann07}.
Neuromodulators such as dopamine have been identified as the necessary third factor for consolidation
\citep{Bailey00,Reymann07,Redondo11,Lisman17}.
Indeed, modern computational models of synaptic consolidation
take into account the effect of neuromodulators \citep{Clopath08,Ziegler15}
in a framework reminiscent of the three-factor rule defined by Eqs.
(\ref{eligibility-1}) and (\ref{three-factor-2}) above.
However, there are two noteworthy differences.
First, in contrast to reward-based learning,
the decay time $\tau_e$ of the synaptic tag $e_{ij}$
is in the range of one hour rather than one second,
consistent with
slice experiments \citep{Frey97}
as well as with behavioral experiments \citep{Moncada07}.
Second, in slices, the measured synaptic weights $w_{ij}$ are increased a few minutes after the end of the induction protocol and decay back with the time course of the synaptic tag whereas in the simplest implementation of the three-factor rule framework as formulated in Eqs.
(\ref{eligibility-1}) and (\ref{three-factor-2}) the visible weight is only updated at the moment when the third factor is present.
However, slightly more involved models where the visible weight depends on both the synaptic tag variable and the long-term stable weight \citep{Clopath08,Ziegler15}
correctly account for the time course of the measured synaptic weights in consolidation experiments \citep{Frey97,Reymann07,Redondo11}.

In summary, the neoHebbian three-factor rule framework has a wide range of applicability. It is well established in the context of synaptic consolidation where the duration of the flag ('synaptic tag') extracted from slice experiments \citep{Frey97} is in the range of one hour,
consistent with fear conditioning experiments \citep{Moncada07}. This time scale is significantly longer than what is needed for
behavioral learning of elementary actions or for memorizing surprising events.
Theoreticians therefore  hypothesized that a process analogous to setting a tag ('eligibility trace')
must also exist on the time scale of one second. The next section discusses some recent experimental evidence supporting this theoretical prediction.

\begin{figure}
    \hbox{ {\bf A}\hspace{7cm} {\bf B}}
\hspace{5mm}
  \includegraphics[width=6.5cm]{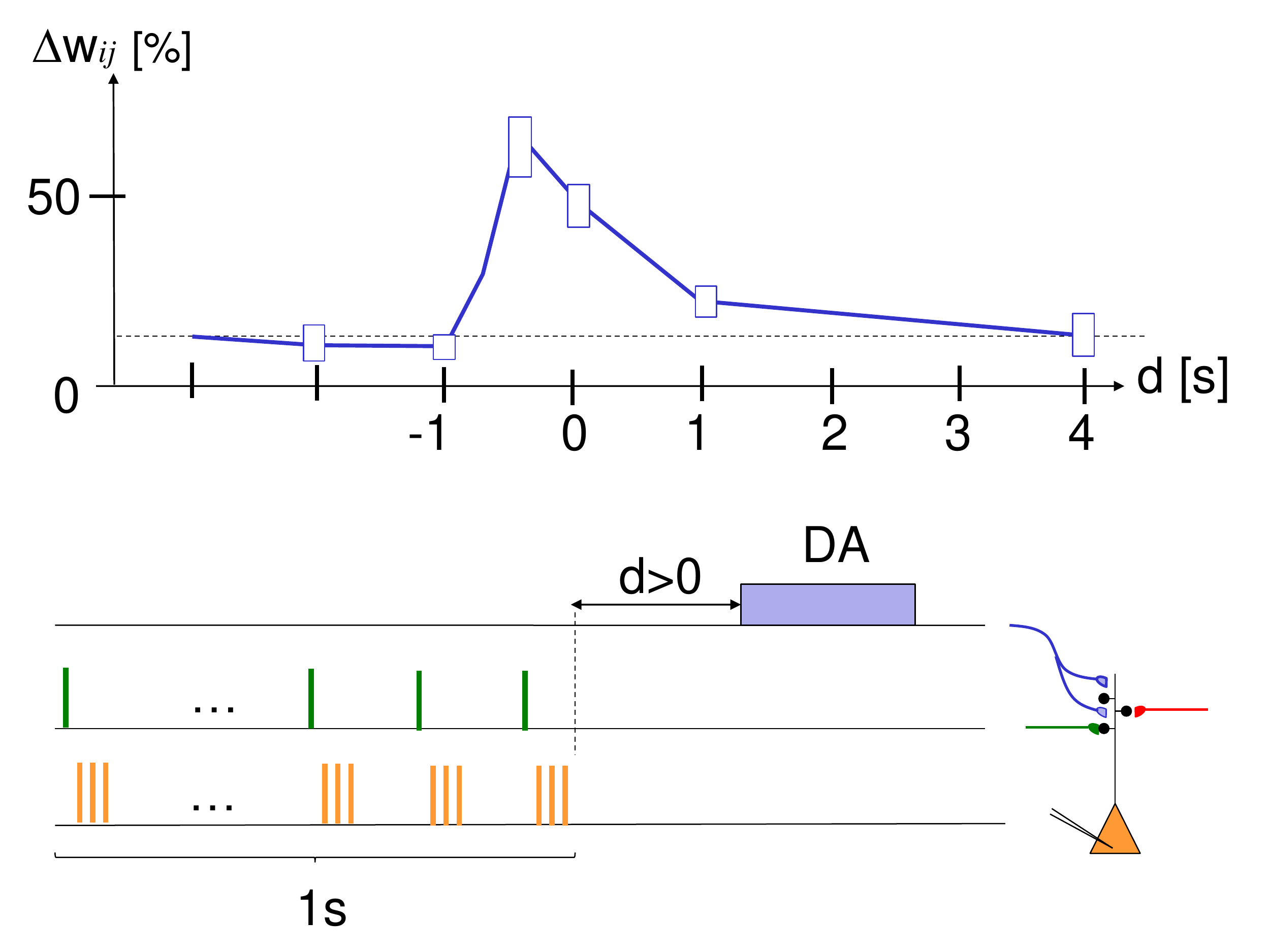}
  \hspace{5mm}
  \includegraphics[width=6.5cm]{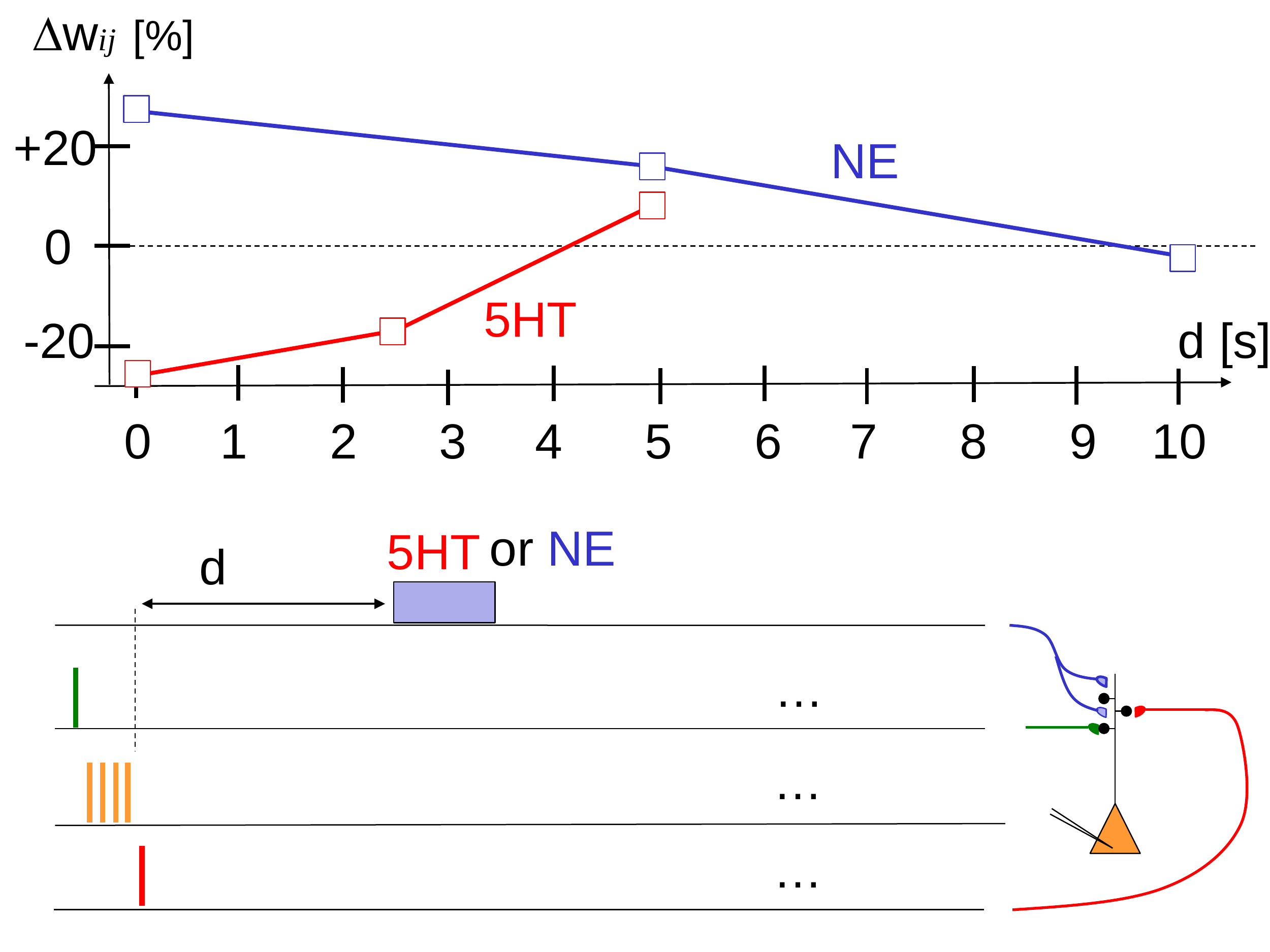}

      \hbox{ {\bf C}\hspace{7cm} {\bf D}}
\hspace{5mm}
  \includegraphics[width=6.5cm]{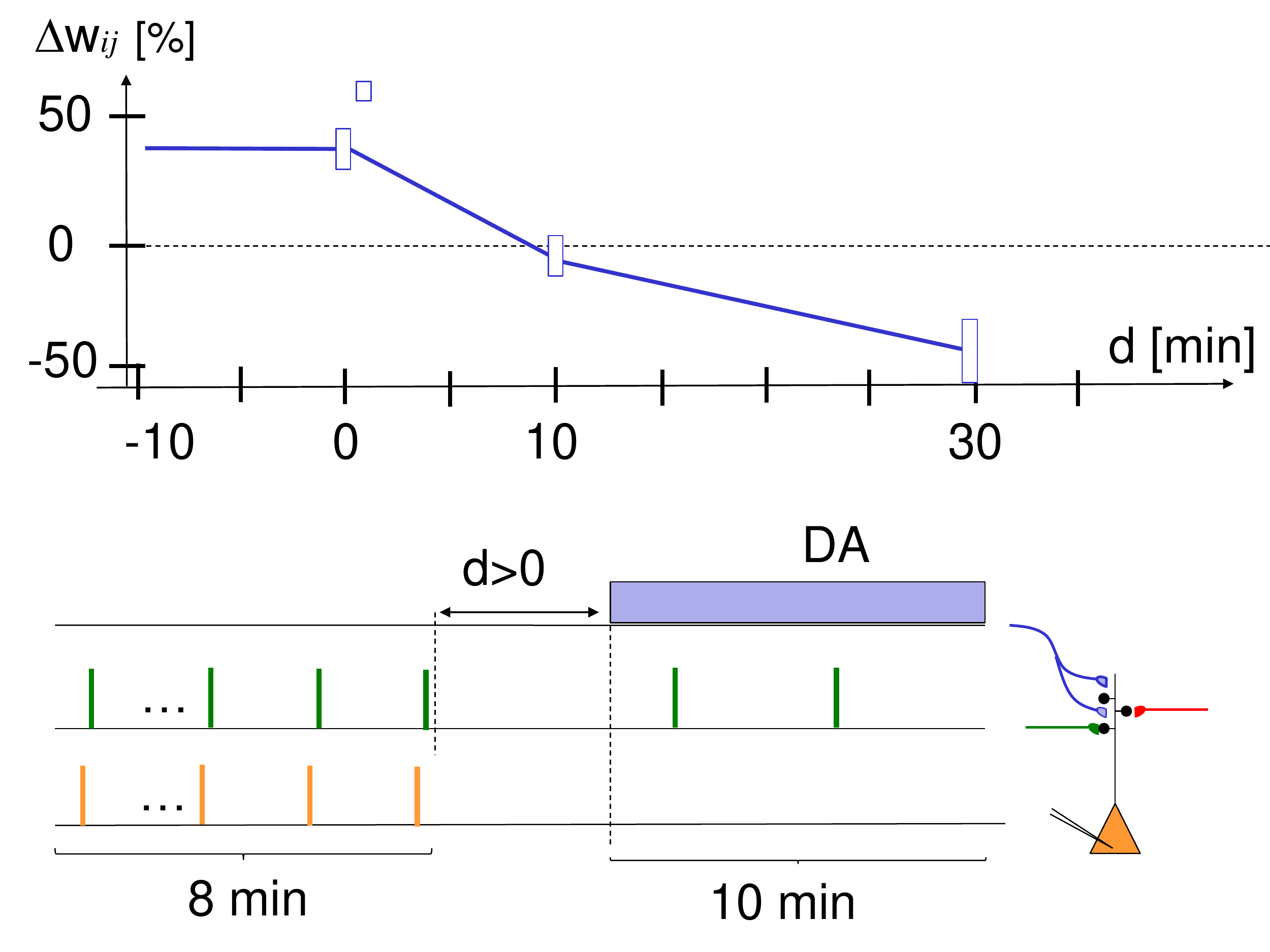}
  \hspace{5mm}
  \includegraphics[width=6.5cm]{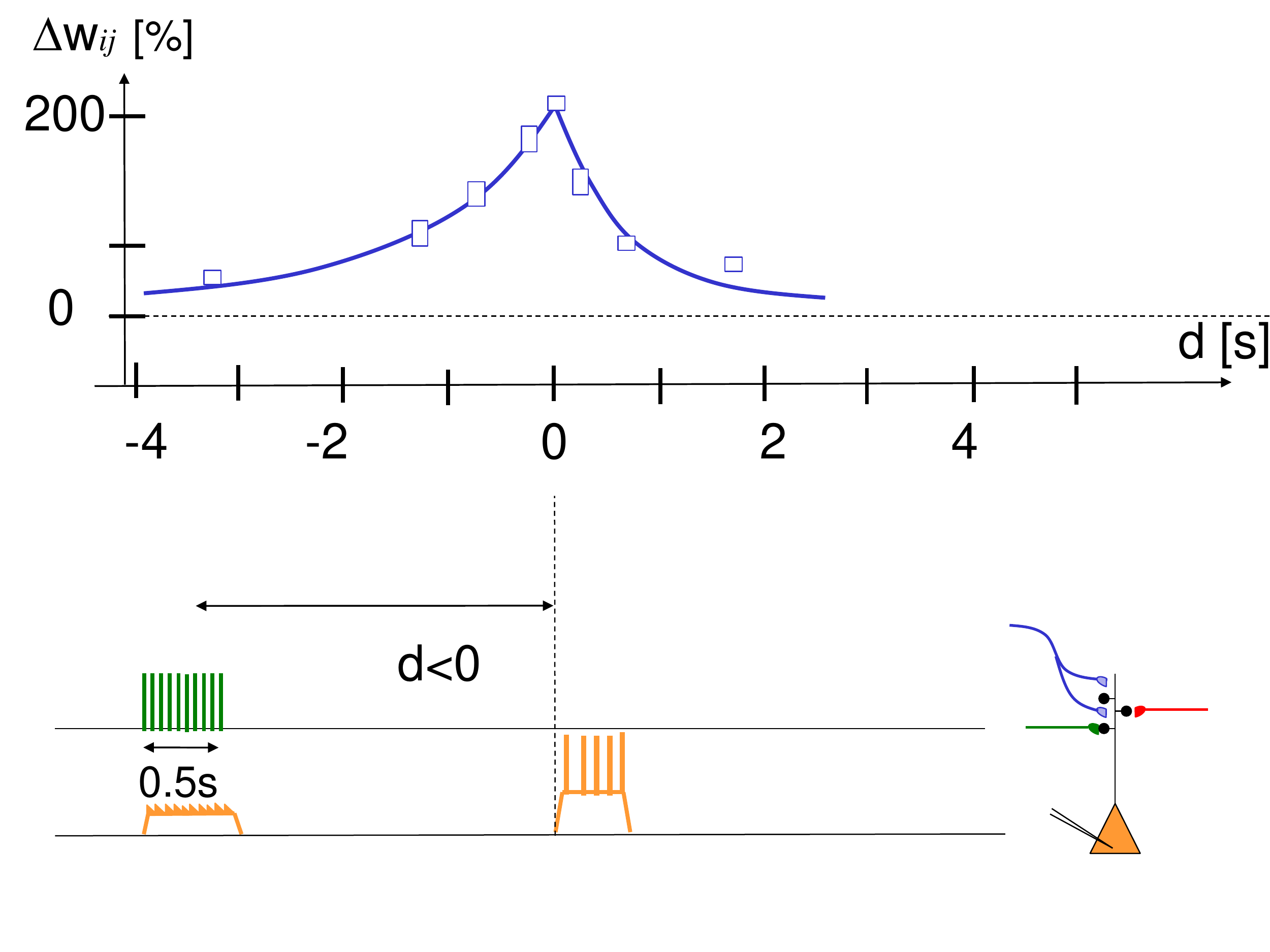}
\caption{\label{fig-2}
    Experimental support for synaptic eligibility traces. Fractional weight change (vertical axis) as a function of delay $d$ of third factor (horizontal axis) for various protocols (schematically indicated at the bottom of each panel).
    {\bf A.}
    In striatum medial spiny cells, stimulation of presynaptic glutamatergic fibers (green)
    followed by three postsynaptic action potentials (STDP with pre-post-post-post at +10ms) repeated 10 times at 10Hz yields LTP if dopamine fibers are stimulated during the presentation ($d<0$) or shortly afterward ($d=0$s or $d=1$s)
    but not if dopamine is given with a delay $d=4$s; redrawn after Fig. 1 of \citep{Yagishita14}, with delay $d$ defined as time since end of STDP protocol.
   {\bf B.}
    In cortical layer 2/3 pyramidal cells, stimulation of two independent presynaptic pathways (green and red) from layer 4 
    to layer 2/3 by a single pulse combined with a burst  of four postsynaptic spikes (orange).
    If the pre-before-post stimulation was combined with a pulse of norepinephrine (NE) receptor agonist isoproterenol with a delay of 0 or 5s, the protocol gave LTP (blue trace). If the post-before-pre stimulation was combined with a pulse of srotonin (5-HT)
    of a delay of 0 or 2.5s, the protocol gave LTD (red trace);
    redrawn after Fig. 6 of \citep{He15}.
    {\bf C.}
    In hippocampus CA1, a post-before-pre ($\Delta t$=-20ms) induction protocol yields LTP if dopamine is present during induction or given with a delay $d$ of zero or one minute, but yields LTD if dopamine is absent or given with a delay of 30min;
    redrawn after Figs 1F, 3C and 2B (square data point at delay of 1min) of  \citep{Brzosko15}.
{\bf D.}
    In hippocampus CA1, 10 extracellular stimuli of presynaptic fibers at 20Hz cause depolarization of the postsynaptic potential. The timing of a  complex spike (calcium plateau potential) triggered by current injection (during 300ms) after a delay $d$, is crucial for the amount of LTP.
    If we interpret presynaptic spike arrival as the first, and postsynaptic depolarization as the second factor,
    the complex spike could be associated with a third factor;
    redrawn after Fig.3 of \citep{Bittner17}.
    Hight of boxes gives a very rough estimate of standard deviation - see original papers and figures for details.
  }
\end{figure}

\section{Experimental evidence for eligibility traces}

Recent experimental evidence for eligibility traces in
striatum \citep{Yagishita14},
cortex \citep{He15},
and hippocampus \citep{Brzosko15,Brzosko17,Bittner17}
is reviewed in the following three subsections.

\subsection{Eligibility traces in dendritic spines of medial spiny striatal neurons in nucleus accumbens}
In their elegant imaging experiment of dendritic spines of nucleus accumbens neurons,
Yagishita et al. (2014)  mimicked presynaptic spike arrival by glutamate uncaging (presynaptic factor), paired it with three postsynaptic spikes immediately afterward (postsynaptic factor), repeated this STDP-like pre-before-post sequence ten times, and combined it with optogenetic stimulation of dopamine fibers (3rd factor) at various delays \citep{Yagishita14}.
The ten repetitions of the pre-before-post sequence at 10 Hz took about one second while
stimulation of dopaminergic fibers (10 dopamine pulses at 30Hz) 
projecting
from the ventral tegmental area (VTA ) to nucleus accumbens 
took about 0.3s.
  In their paper, dopamine 
  was counted as delayed by one second if the dopamine stimulation started immediately after the end of the one-second long induction period (delay = difference in switch-on time of STDP and dopamine), but for consistency with other data we define the delay $d$ here as the time passed since the end of the
  STDP protocol.
  After 15 complete trials the spine volume, an indicator of synaptic strength \citep{Matsuzaki01}, was measured
  and compared with the spine volume before the induction protocol.
  The authors found that 
dopamine promoted
spine enlargement only if phasic dopamine was given
in a narrow time window during or immediately after the 1s-long
STDP protocol; cf. Fig. \ref{fig-2}A.


The maximum enlargement of spines occurred
if the dopamine signal started during the STDP protocol ($d=-0.4$s), but even at a delay of $d= 1$s LTP was still visible.
Giving dopamine too early ($d=-2$s) or too late ($d=+4$s) had no effect. Spine enlargement corresponded to an increase in the amplitude of excitatory postsynaptic currents indicating that the synaptic weight was indeed strengthened after the protocol \citep{Yagishita14}.
Thus, we can summarize that we have in the {\em striatum a three-factor learning rule for the induction of LTP where the decay of the eligibility trace occurs on a time scale of 1s};
cf. Fig. \ref{fig-2}A.

To arrive at these results, Yagishita et al. (2014) concentrated on 
medial spinal neurons 
in the nucleus accumbens core, a part of the ventral striatum of the basal ganglia.
Functionally, striatum is a particularly interesting candidate for reinforcement learning
\citep{Brown95,Schultz98,Doya00a,Arleo00,Daw05} for several reasons.
First, striatum receives highly processed sensory information 
from neocortex and hippocampus through glutamatergic synapses \citep{Mink96,Middleton00,Haber06}.
Second, striatum also receives dopamine
input 
associated with reward processing
\citep{Schultz98}.
Third, striatum is, together with frontal cortex,
involved in selection of motor action programs \citep{Mink96,Seo12}.

On the molecular level, the striatal three-factor plasticity  depended on NMDA, CaMKII, protein synthesis, and dopamine D1 receptors \citep{Yagishita14}. 
CaMKII increases were found to be localized in the spine and to have
roughly the same time course as the  critical window for phasic dopamine suggesting that
CaMKII could be involved in the 'synaptic flag' triggered by the STDP-like induction protocol
while protein kinase A (PKA) was found to have an unspecific cell-wide distribution 
suggesting an interpretation of PKA as a molecule linked to the
dopamine-triggered third factor
\citep{Yagishita14}.

\subsection{Two distinct eligibility traces for LTP and LTD in cortical synapses}
In a recent experiment of He et al. (2015),
layer 2/3 pyramidal cells  in slices from prefrontal or visual cortex were stimulated by an STDP protocol, either pre-before-post for LTP induction or post-before-pre for LTD induction. A neuromodulator was applied with a delay after a single STDP sequence before the whole protocol was repeated;
cf. Fig. \ref{fig-2}B.
Neuromodulators, either norepinephrine (NE), serotonin (5-HT), dopamine (DA), or acetylcholine (ACh) were
ejected from a pipette for 10 seconds or from endogenous fibers (using optogenetics) for one second
\citep{He15}.
It was found that NE was necessary for LTP whereas
5-HT was necessary for LTD.
DA or ACh agonists had no effect in visual cortex but
DA had a positive effect on LTP induction in frontal cortex \citep{He15}.

For the  STDP protocol, He et al. (2015) used
extracellular stimulation of two presynaptic pathways from layer 4 to layer 2/3
(presynaptic factor) combined with a burst of 4 postsynaptic action potentials (postsynaptic factor), either pre-before-post or post-before-pre.
In a  first variant of the experiment, the STDP stimulation was
repeated 200 times at 10 Hz corresponding to a total stimulation time of 20s before the NE or 5-HT was given.
In a second variant, instead of an STDP protocol, they paired presynaptic stimulation
(first factor) with postsynaptic depolarization (second factor)
to -10mV to induce LTP, or to -40mV to induce LTD.
With both protocols it was found that LTP can be induced if the neuromodulator
NE (third factor) arrived with a delay of 5 seconds or less after the LTP protocol, but not 10 seconds.
LTD could be induced if  5-HT (third factor) arrives with a delay of 2.5 seconds or less after the LTD protocol, but not 5 seconds
\citep{He15}.

A third variant of the experiment involved optogenetic stimulation of the
noradrenaline, dopamine, or serotonin pathway
by repeated light pulses during 1s applied immediately, or a few seconds,  after a minimal
STDP protocol consisting 
of a single presynaptic and four postsynaptic pulses (either pre-before-post or post-before-pre),
 a protocol that is physiologically more plausible.
The minimal sequence of STDP pairing and neuromodulation was repeated 40 times
at intervals of 20s.
Results with optogenetic stimulation
were consistent with those mentioned above and showed in addition that
application of NE or 5-HT immediately before the STDP stimulus did not induce LTP or LTD.
{\em Overall these results indicate that in visual and frontal cortex, pre-before-post pairing leaves an eligibility trace that decays over 5-10 seconds and that can be converted into LTP by the neuromodulator noradrenaline.
  Similarly, post-before-pre pairing leaves a shorter eligibility trace that decays over 3 seconds and can be converted into LTD by the neuromodulator serotonin};
cf. Fig. \ref{fig-2}B.

Functionally, a theoretical model in the same paper
\citep{He15}
showed that the measured three-factor learning rules
with two separate eligibility traces  stabilized and prolonged  network activity so as to allow 'event prediction'.
The authors hypothesized that these three-factor rules were related to reward-based learning
in cortex such as perceptual learning in monkey \citep{Schoups01} or mice \citep{Poort15} or reward prediction \citep{Shuler06}.
The relation to surprise was not discussed but might be a direction for further explorations.

Molecularly, the transformation of the Hebbian pre-before-post eligibility trace into LTP  involves beta adrenergic receptors and intracellular cyclic AMP whereas
the transformation of the post-pre eligibility trace into LTD involves
the 5-HT$_{2c}$  receptor \citep{He15}. Both receptors
are anchored at the postsynaptic density consistent with a role in the transformation of an eligibility trace into actual weight changes \citep{He15}.

\subsection{Eligibility traces in hippocampus}
Two experimental groups 
studied eligibility traces in CA1 hippocampal neurons using complementary approaches.
In the studies of Brzosko et al. (2015,2017),
CA1 neurons in hippocampal slices were stimulated during about 8 minutes in an
STDP protocol involving 100 repetitions (at 0.2Hz) of pairs of one extracellularly delivered  presynaptic stimulation pulse (presynaptic factor) and one
postsynaptic action potential (postsynaptic factor) \citep{Brzosko15}.
Repeated pre-before-post with a relative timing  +10 ms gave LTP (in the presence of natural endogenous dopamine)
whereas post-before-pre (-20ms) gave LTD.
However, with additional dopamine (third factor) in the bathing solution,
post-before-pre at -20ms gave LTP \citep{Zhang09}.
Similarly, an STDP protocol with post-before-pre at -10ms resulted in LTP when
endogenous dopamine was present, but in LTD when dopamine was blocked \citep{Brzosko15}.
Thus dopamine broadens the STDP window for LTP into the post-before-pre regime \citep{Zhang09,Pawlak10}.
Moreover, in the presence of ACh during the STDP stimulation protocol, pre-before-post at +10ms 
also gave LTD \citep{Brzosko17}. Thus ACh broadens the LTD window.

The crucial experiment of Brzosko et al. (2015)
involved a delay in the dopamine \citep{Brzosko15}.
Brzosko et al.  started to perfuse dopamine either immediately after the end of the post-before-pre (-20ms) induction protocol or with a delay.  Since the dopamine was given for about 10 minutes, it cannot be considered as a phasic signal -- but at least the {\em start} of the dopamine perfusion was delayed. Brzosko et al. found that
the stimulus that would normally have given LTD turned into LTP if the delay of dopamine was
in the range of one minute or less, but not if dopamine started 10 minutes
after the end of the STDP protocol \citep{Brzosko15}.
Note that for the conversion of LTD into LTP, it was important that the synapses were weakly stimulated at low rate while dopamine was present.
Similarly, a prolonged pre-before-post protocol at +10ms in the presence of ACh gave
rise to LTD, but with dopamine given with a delay of less than 1 minute the same protocol gave LTP
\citep{Brzosko17}.
To summarize, {\em in the hippocampus a prolonged post-before-pre protocol
  (or a pre-before-post protocol in the presence of ACh) yields visible LTD,
  but also sets an invisible synaptic flag for LTP. If dopamine is applied with a delay of less than one minute, the synaptic flag is converted into a positive weight change under continued weak presynaptic stimulation};
cf. Fig. \ref{fig-2}C.

Molecularly, the
conversion of LTD into LTP after repeated stimulation of pre-before-post pulse pairings depended on NMDA receptors and on the cyclic adenosine monophosphate (cAMP) - PKA signaling cascade
\citep{Brzosko15}.
The source of
dopamine could be in the Locus Coeruleus
which would make a link to arousal and novelty \citep{Takeuchi16} 
or from other  dopamine nuclei linked to reward \citep{Schultz98}.
Since the time scale of the synaptic flag reported in \citep{Brzosko15,Brzosko17}
was in the range of minutes, the process studied by Brzosko et al. could be
related to synaptic consolidation \citep{Frey97,Reymann07,Redondo11,Lisman17}
rather than
eligibility traces in reinforcement learning where shorter time constants are needed
\citep{Izhikevich07,Legenstein08b,Fremaux10,Fremaux13}.
The computational study in Brzosko et al. (2017) used
an eligibility trace with a time constant of 2s and showed that dopamine as a reward signal
induced learning of reward location while ACh during exploration enabled a fast relearning after a shift of the reward location \citep{Brzosko17}.

The second study 
combined in vivo with in vitro data \citep{Bittner17}.
From in vivo studies it has been known that
CA1 neurons in mouse hippocampus can develop a novel, reliable,
and rather broadly tuned, place field in a single trial
under the influence of a 'calcium plateau potential'
\citep{Bittner15}, visible as a complex spike at the soma.
Moreover, an artificially induced complex spike
was sufficient to induce
such a novel place field in vivo
\citep{Bittner15,Bittner17}.

In additional slice experiments, several input fibers from CA3 to CA1 neurons were stimulated by 10 pulses from
an extracellular electrode during one second. The resulting nearly synchronous inputs at, probably,  multiple synapses
caused a total EPSP that was about 10mV above baseline at the  soma, and potentially somewhat larger in the dendrite, but did not cause somatic spiking of the CA1 neuron.
The stimulated synapses showed LTP if the presynaptic stimulation was paired
with a calcium plateau potential (complex spike) in the postsynaptic neuron.
LTP occurred, even if the presynaptic stimulation stopped one or two seconds before the start of the plateau potential
or if the plateau potential started before the presynaptic stimulation
\citep{Bittner17}.
The protocol has a remarkable efficiency since potentiation was around 200$\%$ after only 5 pairings.
Thus,
{\em the joint activation of many synapses sets a flag at the activated synapses which is translated into LTP if a calcium
  plateau potential (complex spike) occurs a few seconds before or after the synaptic activation};
cf. Fig. \ref{fig-2}D.
Molecularly, the plasticity processes implied NMDA receptors and calcium channels \citep{Bittner17}.

Functionally, synaptic plasticity in hippocampus is particularly important  because of the role of hippocampus in
spatial memory \citep{OKeefe78}.
CA1 neurons get input from CA3 neurons which have a narrow place field.
The emergence of a broad place field in CA1 has therefore been interpreted as linking
several CA3 neurons (that cover for example the 50 cm of the spatial trajectory
traversed by the rat before the current location) to a single CA1 cell that
codes for the current location \citep{Bittner17}.
Note that at the  typical running speed of rodents, 50 cm correspond to several seconds of running.
The broad activity of CA1 cells has  therefore been interpreted as a predictive representation of
upcoming events or places \citep{Bittner17}.
What could such an upcoming event be?
For a rodent exploring a T-maze it might for example be important
to develop a more precise spatial representation at the T-junction
than inside one of the long corridors.
With a  broad CA1 place field located at the T-junction,
information about the upcoming bifurcation
could become available several seconds before the animal reaches the junction.

Bittner et al. interpreted their findings as the signature of an unusual form of STDP with a particularly long coincidence window on the behavioral time scale
\citep{Bittner17}.
Given that the time span of several seconds between presynaptic stimulation and postsynaptic complex spike is
outside the range of a potential causal relation between input and output,
they classified the plasticity rule as non-Hebbian because the presynaptic neurons do not participate in firing the postsynaptic one \citep{Bittner17}.
As an alternative view, we propose to classify the findings of Bittner et al. as
the signature of an eligibility trace that was left by the joint occurrence of a presynaptic spike arriving from CA3
(presynaptic factor)
and a subthreshold depolarization at the location of the synapse in the postsynaptic CA1 neuron
(postsynaptic factor); cf. Fig. \ref{fig-2}D.
In this view, the setting of the synaptic flag is caused by a 'Hebbian'-type induction,
except that on the postsynaptic side there are no spikes but just  depolarization,
consistent with the role of depolarization as a postsynaptic factor
\citep{Artola93,Ngezahayo00,Sjostrom01,Clopath10}.
In this view, the findings of Bittner et al. suggest that the synaptic flag
set by the induction protocol leaves an eligibility trace which decays over 2s.
If a plateau potential (related to the third factor)
is generated during these two seconds, the eligibility trace caused by the induction protocol is transformed into a
measurable change of the synaptic weight. 
The third factor $M_{3rd}(t)$ in Eq. (\ref{three-factor-2}) could correspond to the complex spike,
filtered with a time constant of about one second.
Importantly,  plateau potentials
can be considered as neuron-wide signals \citep{Bittner15} triggered by surprising novel or rewarding events
\citep{Bittner17}.
In this view, the results of Bittner et al. are consistent with the framework of neoHebbian
three-factor learning rules.
If the plateau potentials are indeed linked to surprising events, the three-factor rule framework predicts that
in vivo many neurons in CA1 receive such a  third input as a broadcast-like signal. However, only those neurons that also get, at the same time, sufficiently strong input from CA3 might develop the visible plateau potential \citep{Bittner15}.

The main difference between the two alternative views is that, in the model discussed in Bittner et al. (2017), {\em each activated synapse} is marked by an  eligibility trace (which is independent of the state of the postsynaptic neuron) whereas in the view of the three-factor rule, the eligibility trace is set only if the presynaptic activation coincides with a strong depolarization of the postsynaptic membrane. Thus, in the model of Bittner et al. the eligibility trace is set by the presynaptic factor alone whereas in the three-factor rule description it is set by the combination of pre- and postsynaptic factors.
The two models can be distinguished in future experiments where
either the postsynaptic voltage is controlled during presynaptic stimulation or where the number of simultaneously stimulated input fibers is minimized. The prediction of the three-factor rule is that
spike arrival at a single synapse, or spike arrival in conjunction with a very small depolarization of less than 2 mV above rest,
is not sufficient to set an eligibility trace. Therefore, LTP will not occur in these cases even if a
calcium plateau potential occurs one second later.



\section{Discussion and Conclusion}



\subsection{Policy gradient versus TD-learning}

Algorithmic models of TD-learning
with discrete states and in discrete time
do not need eligibility traces that extend beyond one time step
\citep{Sutton98}.
In a scenario where the only reward is given in a target state that is several action steps away
from the initial state, reward information shifts, over multiple trials,
from the target state backwards, even if the one-step eligibility trace
connects only one state to the next \citep{Sutton98}.
Nevertheless,
 extended eligibility traces across multiple time steps
are considered  convenient heuristic tools
to speed up learning
in temporal difference algorithms such
as $TD(\lambda)$ or SARSA$(\lambda)$
\citep{Singh96,Sutton98}.

In policy gradient methods \citep{Williams92} as well as
in continuous space-time TD-learning \citep{Doya00,Fremaux13} eligibility traces
appear naturally in the formulation of the problem of reward maximization.
Importantly, a large class of TD-learning and policy gradient methods can be formulated
as three-factor rules for spiking neurons 
where the third factor is defined as reward minus expected reward \citep{Fremaux16}.
In policy gradient methods and related three-factor rules, expected reward is calculated
as a running average of the reward
\citep{Fremaux10} or fixed to zero by choice of reward schedule
\citep{Florian07,Legenstein08b}.
In TD-learning the expected reward in a given time step is defined
as the difference of the value of the current state and that of the next state
\citep{Sutton98}.
In the most recent large-scale applications of reinforcement learning
the expected immediate reward in policy gradient is calculated by a TD-algorithm for state-dependent value estimation
\citep{Greensmith04,Mnih16}.
An excellent modern summary of Reinforcement Learning Algorithms and their historical predecessors can be found in \citep{Sutton18}.

\subsection{Specificity}
If phasic neuromodulator signals are broadcasted over large areas of the brain, the question arises
whether  synaptic plasticity can still be selective.
In the framework of three-factor rules, specificity is inherited
from the synaptic flags which are set by the combination of presynaptic spike arrival {\em and}
an elevated postsynaptic voltage at the location of the synapses.
The requirement is met only for a small subset of synapses,
because presynaptic alone or postsynaptic activity alone
are not sufficient; cf. Fig. \ref{fig-1}B.
Furthermore, among all the flagged synapses only those that show, over many trials,
a correlation with the reward signal will be consistently reinforced
\citep{Legenstein08b,Loewenstein06}.
Specificity can further be enhanced by an attentional feedback mechanism
\citep{Roelfsema05,Roelfsema10} that restricts the number of eligible synapses
to the 'interesting' ones, likely to be involved in the task. Such an attentional
gating signal acts as an additional factor and turns the three-factor 
into a four-factor learning rule \citep{Rombouts15}.

\subsection{Mapping to Neuromodulators}
The third factor is likely to be related to neuromodulators, but 
from the perspective of a theoretician there is no need
to assign one neuromodulator to surprise and another one to
reward. Indeed, the theoretical framework also works
if each neuromodulator codes for a different combination
of variables such as surprise, novelty or reward,
just as we can use different coordinate systems
to describe the same physical system
\citep{Fremaux16}.
Thus, whether dopamine is purely reward related or also
novelty related
\citep{Ljunberg92,Schultz98,Redgrave06}
is not critical for the development of three-factor learning rules
as long as dimensions relating to
novelty, surprise, and reward are all covered by the set of neuromodulators.

Complexity in biology is increased by the fact that
dopamine neurons projecting from the VTA
to the striatum can have separate circuits and functions changing 
from reward in ventral striatum to novelty in the the tail of striatum
\citep{Menegas17}.
Similarly,
dopaminergic fibers starting in  the VTA
can have a different function than those starting in Locus Coeruleus
\citep{Takeuchi16}.
The framework of three-factor rules is general enough to allow for these, and many other,
variations.

\subsection{Alternatives to eligibility traces for bridging the gap between the behavioral and neuronal timescales}

From a theoretical point of view, there is nothing -- apart conceptual
elegance -- to favor eligibility traces over alternative neuronal mechanisms to
associate events that are separated by a second or more.
For example, 
 memory traces hidden in the rich firing activity patterns
of a recurrent network 
\citep{Maass02,Jaeger04,Buonomano09,Sussillo09} or short-term
synaptic plasticity in recurrent networks \citep{Mongillo08}
could be involved in
learning behavioral tasks with delayed feedback.
In some  models, neuronal, rather than synaptic, activity traces have been
involved in learning a delayed paired-associate task
\citep{Brea16a}
and a combination of
synaptic eligibity traces with prolonged single-neuron activity has been used for 
learning on behavioral time scales \citep{Rombouts15}.
The empirical studies reviewed
here support the idea that the brain makes use of the elegant solution
with synaptic eligibility traces and three-factor learning rules, but do not exclude that other mechanisms work in parallel.

\subsection{The paradoxical nature of predictions in computational neuroscience}
If a neuroscientist thinks of a  theoretical model, he often 
imagines a couple of assumptions at the beginning, a set of results derived from simulations
or mathematical analysis, and ideally a few novel predictions - but is this the way modeling works?
There are at least
two types of predictions in computational neuroscience, detailed predictions and conceptual predictions. Well-known examples of detailed predictions have been generated from 
variants of multi-channel biophysical Hodgkin-Huxley type
\citep{Hodgkin52}
models such as: 'if channel X is blocked then we predict that ... ' where X is a channel with known dynamics and predictions include depolarization, hyperpolarization, action potential firing, action potential backpropagation or failure thereof.  All of these are useful predictions readily translated to and tested in experiments.

Conceptual predictions derived from abstract conceptual models are potentially more interesting, but more
difficult to formulate.
Conceptual models develop ideas and form our thinking of
how a specific neuronal system could work to solve a behavioral task such as working memory
\citep{Mongillo08},
action selection and decision making \citep{Sutton98},
long-term stability of memories \citep{Lisman85,Crick84,Fusi05},
memory formation and memory recall \citep{Willshaw69,Hopfield82}.
Paradoxically these models often make no detailed predictions in the sense indicated above.
Rather, in these and other conceptual theories, the most relevant model features
are formulated as {\em assumptions} which may be considered, in a loose sense, as playing the role of {\em conceptual predictions}.
To formulate it as a short slogan: 
Assumptions are predictions.
Let us return to the conceptual framework of three-factor rules: the purification of rough ideas
into the role of three factors is the important conceptual work - and part of the assumptions.
Moreover, the specific choice of time constant in the range of one second
for the eligibility trace has been formulated by theoreticians as one of the model assumptions,
rather than as a prediction; cf. the footnotes in section 'Examples and theoretical predictions'. Why is this the case?

Most theoreticians shy away from calling their conceptual modeling
work a 'prediction', because there is no logical
necessity that the brain must work the way they assume in their model
- the brain could have found a less elegant, different,  but nevertheless functional solution to the problem under consideration; see the examples in the previous subsection.
What a good conceptual model in computational neuroscience shows is that there {\em exists} a (nice) solution
that should ideally not be in obvious contradiction with too many known facts.
Importantly, conceptual models necessarily rely on assumptions which in many cases have not (yet) been shown to be true.
The response of referees to modeling work
in experimental journals therefore often is: 'but this has never been shown'.
Indeed, some assumptions may look far-fetched or even in contradiction with known facts:
for example, to come back to eligibility traces,
experiments on  synaptic tagging-and-capture
have shown in the 1990s  that the time scale of a synaptic flag
is in the range of one {\em hour} \citep{Frey97,Reymann07,Redondo11,Lisman17},
whereas the theory of eligibility traces for action learning
needs a synaptic flag on the time scale of one {\em second}.
Did synaptic tagging results imply that three-factor rules for action learning were wrong, because they used the wrong time scale? Or,
on the contrary,
did these experimental results rather imply that a biological machinery for three-factor rules was indeed in place 
which could therefore, for other neuron types and brain areas, be used and re-tuned to a different time scale \citep{Fremaux13}?

As mentioned earlier,
the concepts of eligibility traces and three-factor rules can be traced back to the 1960ies,
from models formulated in words \citep{Crow68},
to firing rate models formulated in discrete time and discrete states
\citep{Klopf72,Sutton81,Barto83,Barto85,Williams92,Schultz98,Sutton98,Bartlett99},
to models with
spikes in a continuous state space and an explicit time scale for eligibility traces
\citep{Xie04,Loewenstein06,Florian07,Izhikevich07,Legenstein08b,Vasilaki09,Fremaux13}.
Despite the mismatch with the known time scale of synaptic tagging in hippocampus (and lack
of experimental support in other brain areas), 
theoreticians persisted, polished their theories, talked at conferences about these models,
until eventually the experimental techniques and the scientific interests of experimentalists were aligned to directly test the assumptions of these theories. In view of the long history of three-factor learning rules, 
the recent elegant experiments \citep{Yagishita14,He15,Brzosko15,Brzosko17,Bittner17}
provide an instructive example of how conceptual theories can influence experimental neuroscience.

\section*{Funding and Acknowledgments}
This project has been funded by the European Research Council (grant agreement no. 268 689, ``MultiRules''),
by the European Union
Horizon 2020 Framework Program  under grant agreement no. 720270 (Human
Brain Project),
and by the Swiss National Science Foundation (no. 200020 165538).

\footnotesize
\bibliographystyle{chicago}

\end{document}